# A new constitutive model for tetragonal energetic single crystals

Benoit Revil-Baudard [*]

Department of Mechanical & Aerospace Engineering, University of Florida/REEF,

Shalimar, FL 32579, USA

**Abstract**

In this paper is presented a new single crystal plasticity model with yielding accounting for the tetragonal symmetry of molecular energetic crystals. This new tetragonal yield criterion has been developed using representation theorems for anisotropic tensor functions. It is defined for any type of loadings and depends on the characteristics of the tetragonal lattice (c/a ratio). It involves three independent plastic anisotropy parameters that can be expressed analytically in terms of the uniaxial yield stresses along four crystallographic directions. Illustration of the capabilities of this model is done for a pentaerythritol tetranitrate (PETN) crystal. Moreover, we present finite-element meso-scale simulations of the response of a polymer bonded explosive for which the behavior of the constituent energetic molecular crystals is described with an elastic/plastic model with yielding governed by the new single-crystal tetragonal criterion. The simulation results provide insights into the role played by the anisotropic crystalline plasticity and interactions between crystals on the thermo-mechanical response under dynamic compression of the PETN-based polymer bonded explosive (PBX) aggregate.

---

[*] Corresponding author: Tel: +1(850) 833-9350; fax +1(850) 833-9366; E-mail: revil @ ufl.edu (Benoit Revil-Baudard).

1. **Introduction**

Molecular crystals are key components of explosive systems. In general, it is thought that explosives may ignite due to the localization of energy and its conversion into heat in small regions termed "hot spots" (e.g. see Bowden and Yoffe, 1952). One of the mechanisms of hot spot formation is plastic dissipation. Specifically, under low velocity impact conditions or weak shocks heating due to yielding and plastic flow precedes ignition in energetic molecular crystals (e.g. for single crystals of PETN, see Krishna Mohan et al (1989) data for impact at 150 to 200 m.s$^{-1}$).

Moreover, the sensitivity to detonation depends on the structure of the molecular crystals and the crystal orientation (e.g. for PETN crystals, see data on the shock initiation sensitivity reported in Dick, 1984; for other crystal structures, see review paper of Handley et al., 2018). Generally, the plastic anisotropy of the constituent energetic crystals is neglected, the plastic response being modeled using the isotropic von Mises model (e.g. see Tran and Udaykumar (2006)). In the studies on explosive assemblies that do account for the anisotropic crystalline plasticity, the classic Schmid law and power-type hardening laws, developed for metallic crystals, are commonly used (e.g. see Clayton and Becker, 2012 for RDX single crystals of orthorhombic class; for polycrystalline cyclotetramethylene-tetranitramine (HMX) of monoclinic symmetry, see Hardin et al, 2014, Wang et al. 2016, Grilli and Koslowski, 2018 for single crystal β-HMX, etc.). As pointed out in the literature, for molecular energetic crystals there is very little experimental information concerning the active slip systems and there is an intense debate concerning the number of slip systems and their critical resolved stresses (e.g. see Hardin et al., 2014). It is generally accepted that prediction of the pre-detonative thermomechanical response of polycrystalline energetic materials remains a formidable challenge and that there is a need for a more appropriate description of the plastic anisotropy of energetic crystals (e.g. see Handley et al., 2018).

The main aim of this work is to develop a yield criterion for energetic molecular crystals of tetragonal symmetry. The mathematical expression of this criterion is derived on the basis of representation theorems for tensor functions. Specifically, it is demonstrated that a quadratic yield criterion that is pressure-insensitive and fulfills the invariance requirements associated with the tetragonal structure should involve only three anisotropy parameters (Section 2). Moreover, it is shown that these anisotropy parameters depend on the lattice (c/a ratio) and can be tied directly to mechanical properties. The capabilities of the developed criterion to describe the influence of the crystal symmetry on the onset of plastic deformation under uniaxial loadings are examined in detail (Section 3). Application of the developed criterion to single crystal PETN and comparison with available data is also provided. An elastic plastic model with yielding described by this new tetragonal criterion is presented (see Section 4) and further used to model the constituent grains in a PBX aggregate. Results of meso-scale FE simulations of the response of the aggregate under dynamic compression are presented in Section 5. We conclude with a summary of the main findings.

## 2. New anisotropic yield criterion for tetragonal energetic crystal

Constitutive models that account for the specific symmetries of tetragonal energetic molecular crystals (e.g. PETN crystal) do not currently exist. To develop an anisotropic yield criterion for tetragonal crystals, we make use of rigorous theorems of representation of tensor functions (see Boehler, 1987; Cazacu et al. (2019)). In this manner, it is ensured that the symmetry invariance properties associated to the lattice structure are automatically satisfied and the exact number of independent anisotropy parameters that ought to be involved in the formulation are deduced.

Indeed, the crystal structure places severe restrictions on the mathematical form of the function defining the onset of plastic deformation. Specifically, the yield function $\Phi(\sigma)$ should be form-invariant under the group of transformations $g$ associated with the respective crystal class, i.e.

$$\Phi(\mathbf{Q}\,\sigma\,\mathbf{Q}^T) = \Phi(\sigma), \tag{1}$$

for any applied stress tensor $\sigma$ and any orthogonal transformation $\mathbf{Q}$ belonging to the symmetry group $g$ ( e.g. see the book of Cazacu et al. (2019)).

Let us denote by $R_{\mathbf{n}}^{\phi}$ the proper rotation (i.e. det $R_{\mathbf{n}}^{\phi} = 1$) through an angle $\phi$ about an axis in the direction of the unit vector $\mathbf{n}$.

Energetic crystals belonging to the tetragonal system have the following symmetry generators: $R_z^{\pi}$ and $R_z^{\pi/2}$, where the $\mathbf{z}$-axis is taken along the $\mathbf{c}$-axis of the lattice depicted in Fig. 1 (e.g. see Hooks et al (2015) for the space group and elastic properties of PETN crystal; for more details on symmetry generators and irreducible invariants for the tetragonal system, see Teodosiu, 1982). Therefore, in order to deduce the mathematical expression of a tetragonal yield function one can use theorems of representations for tensor-valued functions fulfilling rhombic symmetries and further impose the additional constraint of invariance with respect to rotations of $\pi/2$ about the $\mathbf{c}$-axis. Specifically, it has been demonstrated (e.g. see Spencer, 1987) that for a polynomial $P(\sigma)$ to be form-invariant to any transformations belonging to the rhombic system, relative to the Oxyz axes associated to its axes of symmetry it must be expressible as a polynomial in the set of the irreducible invariants:

$$\sigma_{xx},\ \sigma_{yy},\ \sigma_{zz},\ \sigma_{xy}^2,\ \sigma_{yz}^2,\ \sigma_{xz}^2,\ \sigma_{xy}\sigma_{xz}\sigma_{yz}\ , \tag{2}$$

i.e.

$$P(\sigma) = P\left(\sigma_{xx},\ \sigma_{yy},\ \sigma_{zz},\ \sigma_{xy}^2,\ \sigma_{yz}^2,\ \sigma_{xz}^2,\ \sigma_{xy}\sigma_{xz}\sigma_{yz}\right) \tag{3}$$

The requirement of invariance with respect to rotations of $\pi/2$ about the $\mathbf{z}$-axis, means that:

$$P\left(\sigma_{xx},\ \sigma_{yy},\ \sigma_{zz},\ \sigma_{xy}^2,\ \sigma_{yz}^2,\ \sigma_{xz}^2,\ \sigma_{xy}\sigma_{xz}\sigma_{yz}\right)=P\left(\sigma_{yy},\ \sigma_{xx},\ \sigma_{zz},\ \sigma_{xy}^2,\ \sigma_{xz}^2,\ \sigma_{yz}^2,\ \sigma_{xy}\sigma_{xz}\sigma_{yz}\right) \quad (4)$$

Let us denote by **I** the second-order identity tensor ($I_{ij} = \delta_{ij}$ with $i, j = 1\ldots 3$).

Further, imposing the condition of pressure-insensitivity of yielding i.e. $P(\boldsymbol{\sigma}) = P(\boldsymbol{\sigma} + p\mathbf{I})$ for any scalar p we obtain the following general result:

*Relative to the coordinate system Oxyz associated with the symmetry axes of the tetragonal crystal (with the z-axis along the **c**-axis of the tetragonal lattice, see Fig. 1), a quadratic pressure-insensitive yield criterion for a tetragonal crystal should be necessarily of the form:*

$$f(\boldsymbol{\sigma}) = \left\{ \frac{a_1}{6}\left[\left(\sigma_{xx}-\sigma_{zz}\right)^2 + \left(\sigma_{yy}-\sigma_{zz}\right)^2\right] + \frac{a_3}{6}\left(\sigma_{xx}-\sigma_{yy}\right)^2 + a_4\sigma_{xy}^2 + a_5\left(\sigma_{xz}^2 + \sigma_{yz}^2\right) \right\} \quad (5)$$

### 3. Predicted anisotropy in yielding and application of the tetragonal criterion to a PETN crystal

**3.1 Effect of loading orientation on yielding**

Let us denote by $Y_f$ the yield limit under uniaxial loading along a direction **f** of Miller indices [αβγ]. In the Oxyz Cartesian coordinate system associated with the crystallographic axes, at yielding the stress tensor is given by:

$$\boldsymbol{\sigma} = \frac{Y_f}{\alpha^2 + \beta^2 + \gamma^2(c/a)^2}\begin{bmatrix} \alpha^2 & \alpha\beta & \alpha\gamma(c/a) \\ \alpha\beta & \beta^2 & \beta\gamma(c/a) \\ \alpha\gamma(c/a) & \beta\gamma(c/a) & \gamma^2(c/a)^2 \end{bmatrix}_{(\mathbf{e}_x,\mathbf{e}_y,\mathbf{e}_z)} \quad (6)$$

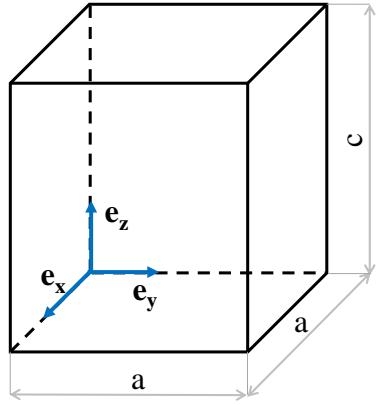

Figure 1. Definition of the reference frame associated with the symmetries of the lattice of the tetragonal crystal

Substituting Eq. (6) into the expression of the yield criterion given by Eq.(5), we obtain the value of $Y_f$, the yield stress under uniaxial loading along the direction **f**.

Let us first examine the predictions for the uniaxial yield stresses along the crystallographic directions of symmetry of the lattice. If $Y_{100}$ denotes the yield stress under uniaxial loading along the [100] (**x**-direction in Fig.1), then according to the single-crystal yield criterion (5) :

$$\frac{Y_{100}}{k} = \sqrt{\frac{6}{a_1 + a_3}} \qquad (7)$$

where $k$ denotes the yield limit in simple shear in the (**x,y**) crystallographic plane ( by symmetry, the yield limit in simple shear in the (yz) crystallographic plane is also equal to $k$).

It is worth noting that the developed criterion automatically respects the symmetries of the tetragonal lattice and as such predicts the same yield limit under uniaxial loading along the $[010], [\bar{1}00], [0\bar{1}0]$ directions, respectively (see Eq. (5)).

On the other hand, $Y_{001}$, the yield stress under uniaxial loading along the **c**-axis (**z**-axis) is:

$$\frac{Y_{001}}{k} = \sqrt{\frac{3}{a_1}}. \tag{8}$$

**Remark**: Comparison between Eq.(7) and Eq.(8) shows that for tetragonal crystals, the yield stress along the [001] direction is different than the yield stress along [100] direction. Unlike the case of cubic single crystals, we have $Y_{001} = Y_{100}$ if and only if $a_1 = a_3$.

It is worth examining the yield stresses along the diagonals of each face of the tetragonal lattice. If $Y_{110}$ denotes the yield stress under uniaxial loading along the [110] crystallographic direction, then the new yield criterion (5) predicts:

$$\frac{Y_{110}}{k} = \sqrt{\frac{12}{a_1 + 3a_4}}. \tag{9}$$

If $Y_{101}$ denotes the yield stress under uniaxial loading along the [101] crystallographic direction, then according to the criterion (5):

$$\frac{Y_{101}}{k} = \sqrt{\frac{6\left(1 + (c/a)^2\right)^2}{a_1 + a_3 - 2a_1(c/a)^2 + 2a_1(c/a)^4 + 6a_5(c/a)^2}}. \tag{10}$$

**Remark**: It is to be noted that the new tetragonal yield criterion predicts the same yield limit under uniaxial loading along the orientations $[011], [\bar{1}0\bar{1}], [0\bar{1}\bar{1}]$, which is a direct consequence of the criterion being form-invariant to the symmetries of the tetragonal structure.

Another crystallographic orientation of interest is the lattice diagonal, i.e. the [111] direction. According to the yield criterion:

$$\frac{Y_{111}}{k} = \sqrt{\frac{3\left(2 + (c/a)^2\right)^2}{a_1 + 3a_4 + 2(c/a)^2(3a_5 - a_1) + a_1(c/a)^4}}. \tag{11}$$

It is also of interest to illustrate that with the developed criterion it is possible to account for the fact that the in-plane anisotropy in uniaxial yield stresses of a tetragonal crystal is dependent on the crystallographic plane considered.

First, let us analyze the effect of the loading orientation on yielding in uniaxial compression in the crystallographic planes of normal <010> (see Fig.2(a)). If $Y_\theta$ is the uniaxial yield stress along a direction at angle $\theta$ to the crystallographic axis [001] ($\theta=0$ represents the direction [001] and $\theta=\pi/2$ is the direction [100]), according to the criterion, we have:

$$\frac{Y_\theta}{Y_{100}} = \sqrt{\frac{a_1+a_3}{\left(\sin^4\theta(a_1+a_3)+2a_1\cos^4\theta+2\sin^2\theta\cos^2\theta(3a_5-2a_1)\right)}} \qquad (12)$$

Note that for $\theta=0$ (i.e. the direction [001]), Eq. (12) gives the ratio between the yield stresses along the **c**-axis and **a**-axis deduced previously (see Eq. (7) and Eq.(8)) while for $\theta=\pi/2$ which corresponds to the direction [100], Eq. (12) reduces to an identity.

Another crystallographic plane of interest is the plane of normal $<1\bar{1}0>$ (see Fig. 2(b)). Let us denote by $Y_{<110>}(\theta)$ the yield stress under uniaxial loading along an arbitrary axis $\mathbf{d}=\sin\theta/\sqrt{2}\,\mathbf{e_x}+\sin\theta/\sqrt{2}\,\mathbf{e_y}+\cos\theta\,\mathbf{e_z}$ in this plane ($\theta=0$ corresponds to the direction [001] and $\theta=\pi/2$ to the direction [110], respectively). According to the criterion,

$$\frac{Y_{<110>}(\theta)}{Y_{100}} = \sqrt{\frac{a_1+a_3}{2\left(\sin^4\theta(a_1+3a_4)/4+a_1\cos^4\theta+\sin^2\theta\cos^2\theta(3a_5-a_1)\right)}} \qquad (13)$$

Note that for $\theta=0$ and $\theta=\pi/2$, Eq. (13) gives the predictions for the ratio between the yield stresses in the [001] and [100] directions, and respectively the ratio between yield stresses in the [110] and [100] directions obtained previously (see Eq. (9)).

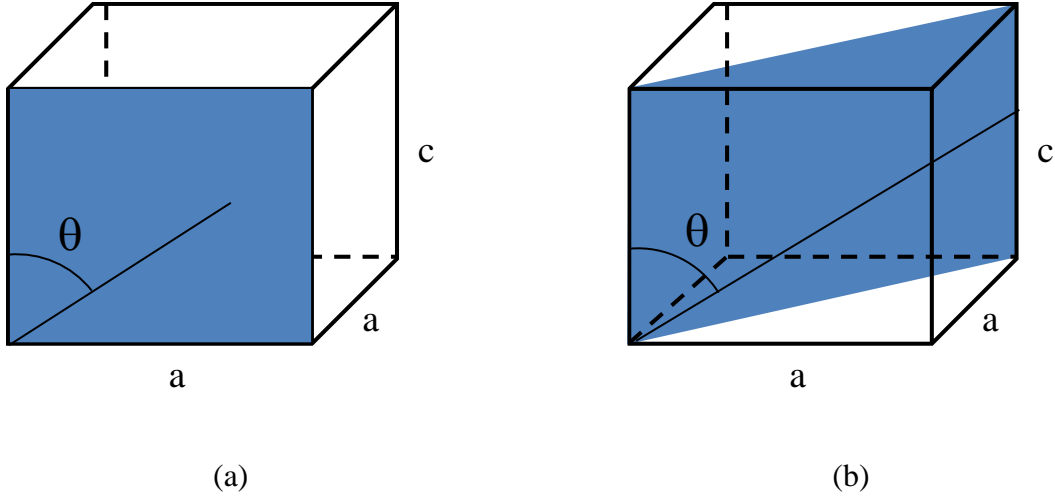

Figure 2. Representation of the crystallographic planes of normal : (a) <010> ; (b) $<1\bar{1}0>$.

## 3.2. Identification procedure for the parameters of the new yield criterion for tetragonal energetic crystals

Note that if we impose that the yield surface passes through the yield point under uniaxial loading along the <100> direction, the tetragonal single crystal yield criterion is expressed as:

$$\frac{6}{a_1+a_3}\left\{\frac{a_1}{6}\left[(\sigma_{xx}-\sigma_{zz})^2+(\sigma_{yy}-\sigma_{zz})^2\right]+\frac{a_3}{6}(\sigma_{xx}-\sigma_{yy})^2+a_4\sigma_{xy}^2+a_5(\sigma_{xz}^2+\sigma_{yz}^2)\right\}=Y_{100}^2 \quad (14)$$

Note that due to the homogeneity of the developed yield function, the yielding response is the same if the coefficients $a_1, a_3, a_4, a_5$ are replaced by $\beta a_1, \beta a_3, \beta a_4, \beta a_5$, with $\beta$ being an arbitrary positive constant (see Eq. (14)). Therefore, without loss of generality one of these parameters, for example $a_1$, can be set equal to unity. From Eq. (7)-(11), it can be easily seen that the remaining coefficients, $a_3, a_4, a_5$ can be expressed in terms of the uniaxial yield stresses along any three different crystallographic directions among the [001], [111], [101] and [110] directions. Indeed, we have:

$$\left(\frac{Y_{001}}{Y_{100}}\right)^2 = \frac{a_1 + a_3}{2a_1}$$

$$\left(\frac{Y_{110}}{Y_{100}}\right)^2 = \frac{2a_1 + 2a_3}{a_1 + 3a_4}$$

$$\left(\frac{Y_{111}}{Y_{100}}\right)^2 = \frac{(a_1 + a_3)\left(2 + (c/a)^2\right)^2}{2a_1 + 6a_4 + 2(c/a)^2(6a_5 - 2a_1) + 2a_1(c/a)^4}$$

$$\left(\frac{Y_{101}}{Y_{100}}\right)^2 = \frac{(a_1 + a_3)\left(1 + (c/a)^2\right)^2}{a_1 + a_3 - 2a_1(c/a)^2 + 2a_1(c/a)^4 + 6a_5(c/a)^2} \tag{15}$$

In summary, the anisotropy parameters involved in the new tetragonal yield criterion can be easily determined from a few uniaxial tests. It is also worth noting that the plastic anisotropy coefficients of a given tetragonal crystal depend on its lattice characteristics (ratio $c/a$ between the lattice dimensions, see Eq. (15)).

### 3.3. Application to a PETN crystal

As an example, the proposed yield criterion for tetragonal crystals is applied to the energetic PETN crystal. Due to difficulties associated with testing energetic crystals the data available is very limited. In the following, we used the data reported in the open literature by Conroy et al. (2007) who reported the uniaxial yield stresses in compression along five crystallographic directions. The data indicates that the PETN crystal is strongly anisotropic, the yield stress along the **c**-axis being almost double the yield stress along any of the **a**-directions; the reported yield stress ratios are: $Y_{001}/Y_{100} = 1.85$, $Y_{101}/Y_{100} = 0.85$, $Y_{111}/Y_{100} = 0.85$ and $Y_{110}/Y_{100} = 1.09$. At ambient temperature (295 K), the PETN crystal has a ratio $c/a$ of 0.7143 (e.g. see Cady and Larson, 1975). Using the reported yield stress ratios $Y_{001}/Y_{100}$, $Y_{111}/Y_{100}$ and $Y_{110}/Y_{100}$ and $a_1 = 1$ in conjunction with Eq. (15), it follows that $a_3 = 5.90$, $a_4 = 3.57$ and $a_5 = 6.08$.

The reported yield ratio $Y_{101}/Y_{100}$ which has not been used for the determination of the parameters involved in the tetragonal yield criterion can serve for validation purposes. Indeed, for the set of values of anisotropy coefficients identified using Eq. (15)$_4$, we predict $Y_{101}/Y_{100} = 0.8$ against $Y_{101}/Y_{100} = 0.85$ reported by Conroy et al. (2007).

Figure 3 shows the projection of the yield locus according to the new yield criterion for the PETN crystal in the biaxial planes $(\sigma_{xx}, \sigma_{yy})$ and $(\sigma_{xx}, \sigma_{zz})$.

It is to be noted that if the plastic anisotropy of the tetragonal energetic crystal is neglected, an isotropic yield criterion such as von Mises yield criterion can be used. To visualize the influence of neglecting the anisotropy of the tetragonal energetic crystal, in Figure 3 is also plotted the yield surface according to the isotropic von Mises yield criterion. It is worth noting that the proposed anisotropic yield criterion is in very good agreement with the data. Obviously, the isotropic yield criterion cannot account for the influence of the loading orientation on the plastic behavior of the PETN crystal. The maximum difference between yielding according to the proposed tetragonal criterion (14) and the isotropic yield criterion of von Mises is for loadings in the $(\sigma_{xx}, \sigma_{zz})$ because von Mises yield criterion cannot account for the strong difference in yield stresses along the [100] (**x**-direction or **a**-direction of the lattice) and [001] (**z**-direction or **c**-direction).

Also, it is important to note that if the von Mises yield criterion is used to describe the PETN single crystal, the plastic dissipation would be the same irrespective of the orientation of the PETN crystal with respect to the applied loading. However, using the proposed yield criterion, which accounts for the intrinsic symmetry of the energetic crystal, the predicted plastic dissipation will depend on the relative orientation of the crystal axes and loading axes.

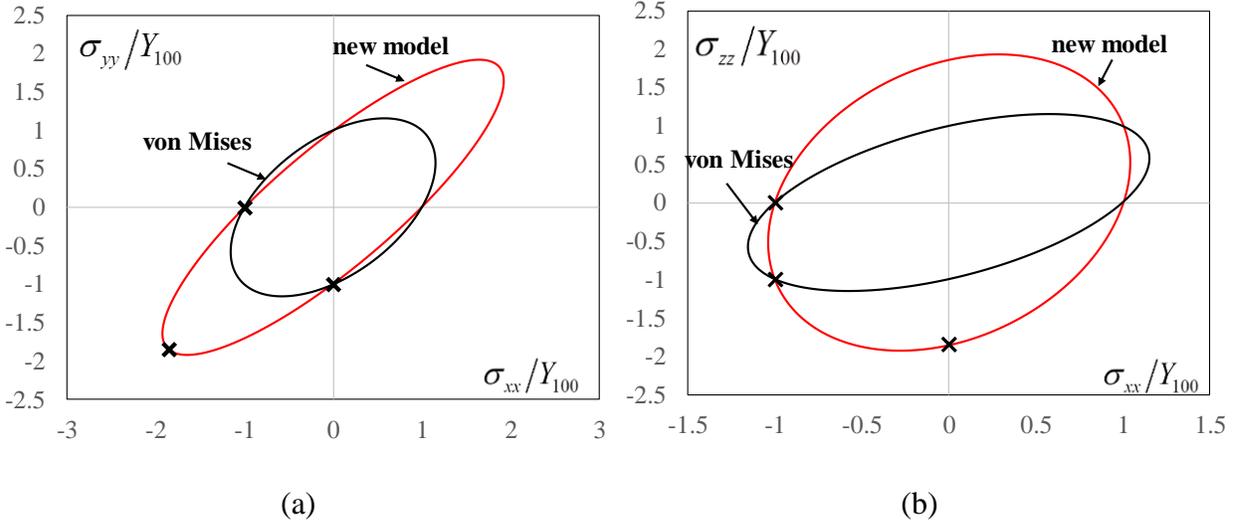

Figure 3. Projections of the yield surface according to the new anisotropic yield criterion accounting for the tetragonal symmetry of the PETN single crystal and the von Mises yield surface (a) in the $(\sigma_{xx}, \sigma_{yy})$ plane (b) in the $(\sigma_{xx}, \sigma_{zz})$ plane. Data from Conroy et al. (2007) are represented by symbols.

For this PETN crystal, in Figure 4 (a) and (b) are shown the predicted variation of the yield stress in uniaxial compression with the orientation $\theta$ in the crystallographic planes of normal <010> and $<1\bar{1}0>$, respectively. In the plane of normal <010>, it is predicted that the ratio $Y_\theta/Y_{100}$ varies between 1.85 which corresponds to the orientation [001] (**c**-axis) and 0.78 for a crystallographic direction close to [506]. In the plane of normal $<1\bar{1}0>$, it is predicted that the ratio $Y_{<110>}(\theta)/Y_{100}$ varies between 1.85 which corresponds to the orientation [001] (i.e. **c**-axis), and 0.8 for an orientation $\theta$ close to $\pi/4$ (see Fig. 2(b)). Note also the good agreement between the model predictions and available data of Conroy et al. (2007).

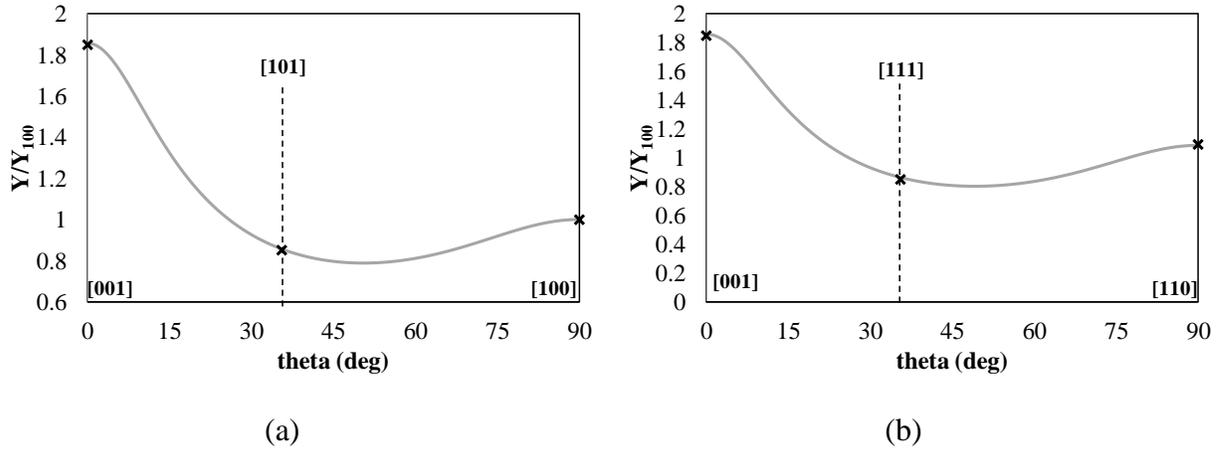

Figure 4. Variation of the normalized yield stress with the loading orientation $\theta$: (a) in the plane of normal $<010>$ (b) in the plane of normal $<1\bar{1}0>$ according to the new yield criterion in comparison with the data for a PETN crystal reported by Conroy et al. (2007) (symbols).

As already discussed, due to the symmetries of the tetragonal crystal, to fully represent the influence of the loading orientation on yielding under uniaxial compression, the predicted response for uniaxial loading along any crystallographic direction contained in the basic stereographic triangle with corners along [001], [110] and [100] should be considered. In Figure 5 is presented in the basic stereographic triangle the predicted uniaxial yield stress normalized by the yield stress along [100] of the PETN single crystal. It is worth noting that for this PETN crystal, the new anisotropic model predicts that the maximum yield stress is along [001], i.e. along the **c**-axis (predicted ratio 1.85) while the lowest yield stress (predicted ratio 0.78) is along an orientation closed to [506] (i.e. absolute minimum value corresponds to the minimum in the yield stresses for uniaxial loading along directions in the plane of normal $<1\bar{1}0>$). Therefore, for the PETN single crystal, the maximum variation in yield stress in uniaxial compression is about 134%.

As previously mentioned, for the prediction of the hot spot sites, i.e. of the local peaks in temperature in an energetic material (either single crystal or energetic aggregate), it is critical to correctly describe the plastic dissipation. The results presented indicate that under uniaxial loading, the local rise of temperature can vary by as much as 134% depending on the orientation of the PETN single crystal (or the grain orientation in an aggregate). On the other hand, if the crystallinity of the energetic system is neglected, irrespective of the loading orientation the yield stress in uniaxial compression will be the same and as a consequence the predicted rise in temperature would be also the same.

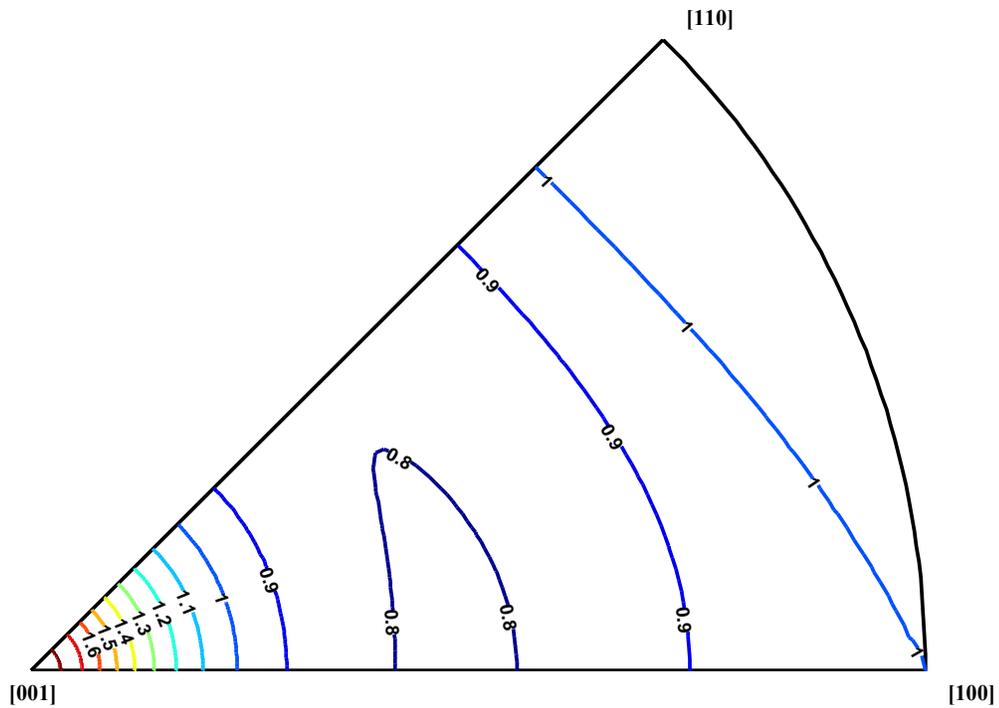

Figure 5. Anisotropy in yield stresses under uniaxial compression for a PETN single crystal according to the proposed yield criterion (Eq.(5)), identified based on the data reported by Conroy et al. (2007). Yield stresses are normalized by the yield stress along [100].

## 4. Constitutive model for the tetragonal single crystal

A constitutive model for energetic tetragonal crystals should account for the influence of the crystal structure on its thermo-mechanical behavior. To this end, an elastic/plastic modeling framework is adopted, with the plastic behavior being described with the proposed tetragonal yield criterion. The governing equations are:

$$\begin{cases} \mathbf{d} = \mathbf{d}^e + \mathbf{d}^p \\ \dot{\boldsymbol{\sigma}} = \mathbf{C}^e : \mathbf{d}^e \\ \varphi(\boldsymbol{\sigma}, \bar{\varepsilon}^p) = f(\boldsymbol{\sigma}) - Y(\bar{\varepsilon}^p, \dot{\bar{\varepsilon}}) \leq 0 \\ \mathbf{d}^p = \dot{\lambda} \dfrac{\partial f}{\partial \boldsymbol{\sigma}} \\ \dot{\bar{\varepsilon}}^p = \dot{\lambda} \end{cases} \quad (16)$$

The Eq. (16)$_1$ expresses that the strain-rate tensor $\mathbf{d}$ is decomposed into an elastic part, $\mathbf{d}^e$, and a plastic part, $\mathbf{d}^p$. For the tetragonal crystal, linear elastic and anisotropic behavior governed by Hooke's law is considered. Namely, there are six independent elastic constants (e.g. see Teodosiu, 1982). In the coordinate system Oxyz associated with the crystallographic axes, the elastic fourth-order tensor $\mathbf{C}^e$ involved in Eq. (16)$_2$ is expressed in Voigt notation as:

$$\mathbf{C}^e = \begin{bmatrix} C_{11} & C_{12} & C_{13} & 0 & 0 & 0 \\ C_{12} & C_{11} & C_{13} & 0 & 0 & 0 \\ C_{13} & C_{13} & C_{33} & 0 & 0 & 0 \\ 0 & 0 & 0 & C_{44} & 0 & 0 \\ 0 & 0 & 0 & 0 & C_{55} & 0 \\ 0 & 0 & 0 & 0 & 0 & C_{55} \end{bmatrix} \quad (17)$$

The elastic response of tetragonal energetic crystals has been thoroughly investigated from both experimental and numerical standpoints (see review paper of Hooks et al (2015)).

The yield surface of the energetic crystal is expressed in the form:

$$\varphi\left(\boldsymbol{\sigma}, \bar{\varepsilon}^{p}\right) = f(\boldsymbol{\sigma}) - Y\left(\bar{\varepsilon}^{p}, \dot{\bar{\varepsilon}}\right) \tag{18}$$

where $f$ is the new anisotropic yield criterion developed (see Eq.(5)) that accounts for the specific symmetries of the energetic crystal, and $Y\left(\bar{\varepsilon}^{p}, \dot{\bar{\varepsilon}}\right)$ is the hardening law of the energetic crystal. Associated plasticity can be considered (see Schmid and Boas, 1950), so the local plastic strain-rate tensor is obtained with the flow rule

$$\mathbf{d}_{i}^{p} = \dot{\lambda} \frac{\partial \bar{\sigma}}{\partial \boldsymbol{\sigma}} \tag{19}$$

with $\dot{\lambda}$ the plastic multiplier and $\bar{\sigma}$ is the effective stress associated with the yield criterion given by Eq. (5) i.e.

$$\bar{\sigma} = \left\{ \frac{a_{1}}{6}\left[\left(\sigma_{xx} - \sigma_{zz}\right)^{2} + \left(\sigma_{yy} - \sigma_{zz}\right)^{2}\right] + \frac{a_{3}}{6}\left(\sigma_{xx} - \sigma_{yy}\right)^{2} + a_{4}\sigma_{xy}^{2} + a_{5}\left(\sigma_{xz}^{2} + \sigma_{yz}^{2}\right) \right\}^{1/2} \tag{20}$$

Note that the hardening law $Y\left(\bar{\varepsilon}^{p}, \dot{\bar{\varepsilon}}\right)$ may depend on the equivalent plastic strain $\bar{\varepsilon}^{p}$ which is defined as the work-conjugate of the effective stress $\bar{\sigma}$ given by Eq. (20) and may be also be function of the strain-rate $\dot{\bar{\varepsilon}}$.

The rise in temperature due to the plastic energy dissipation is calculated as

$$\Delta T = \int_{0}^{t} \frac{\alpha\left(\boldsymbol{\sigma} : \mathbf{d}^{p}\right)}{\rho C} dt \tag{21}$$

where $\alpha$ is the Taylor–Quinney coefficient, $\rho$ the density and C the heat capacity, while $\boldsymbol{\sigma}$ is the stress tensor and $\mathbf{d}^{p}$ the plastic strain-rate tensor, calculated using the new anisotropic plastic model (see Eq. (19)-(20)).

It is important to note that this constitutive model based on the new tetragonal yield criterion is defined for general 3-D loadings. Use of this model for the description of the constitutive behavior of energetic crystals enables the study of the influence of the crystal orientation on its response for complex loadings. To this end, this constitutive model was implemented in a finite-element (FE) framework. Specifically, a user material subroutine (UMAT) was developed and implemented in the FE software Abaqus (see Abaqus, 2014). A fully-implicit backward Euler method was used for the integration of the set of elastic-plastic governing equations given by the system of Eq. (16). In addition to a very good accuracy, the backward Euler method was shown to be unconditionally stable with respect to the size of the strain increment.

**5. Mesoscale F.E. simulation of the temperature rise in an energetic aggregate with constituent PETN grains modeled with the new tetragonal model**

In this section, we present FE. meso-scale simulations of an energetic system consisting of energetic crystals surrounded by a polymer. The primary objective is to study the effect of the crystalline plastic anisotropy on the temperature field in the aggregate. To this end, the constituent crystals are modeled with the new constitutive model with yielding described with tetragonal single crystal criterion developed (see Eq. (16) and Eq.(5) ).

We consider an aggregate composed of 14 PETN crystals surrounded by a polymer binder subjected to dynamic uniaxial compressive loading (see Figure 6) with no lateral displacement on the side of the specimen. The PBX mass composition is 75% PETN ($\rho = 1770$ kg/m$^3$) and 25% Sylgard 182 ($\rho = 1000$ kg/m$^3$). Each crystal has a different shape, size and crystal orientation with respect to the loading axis, the specific orientation of each constituent crystal is given in Figure 6(a).

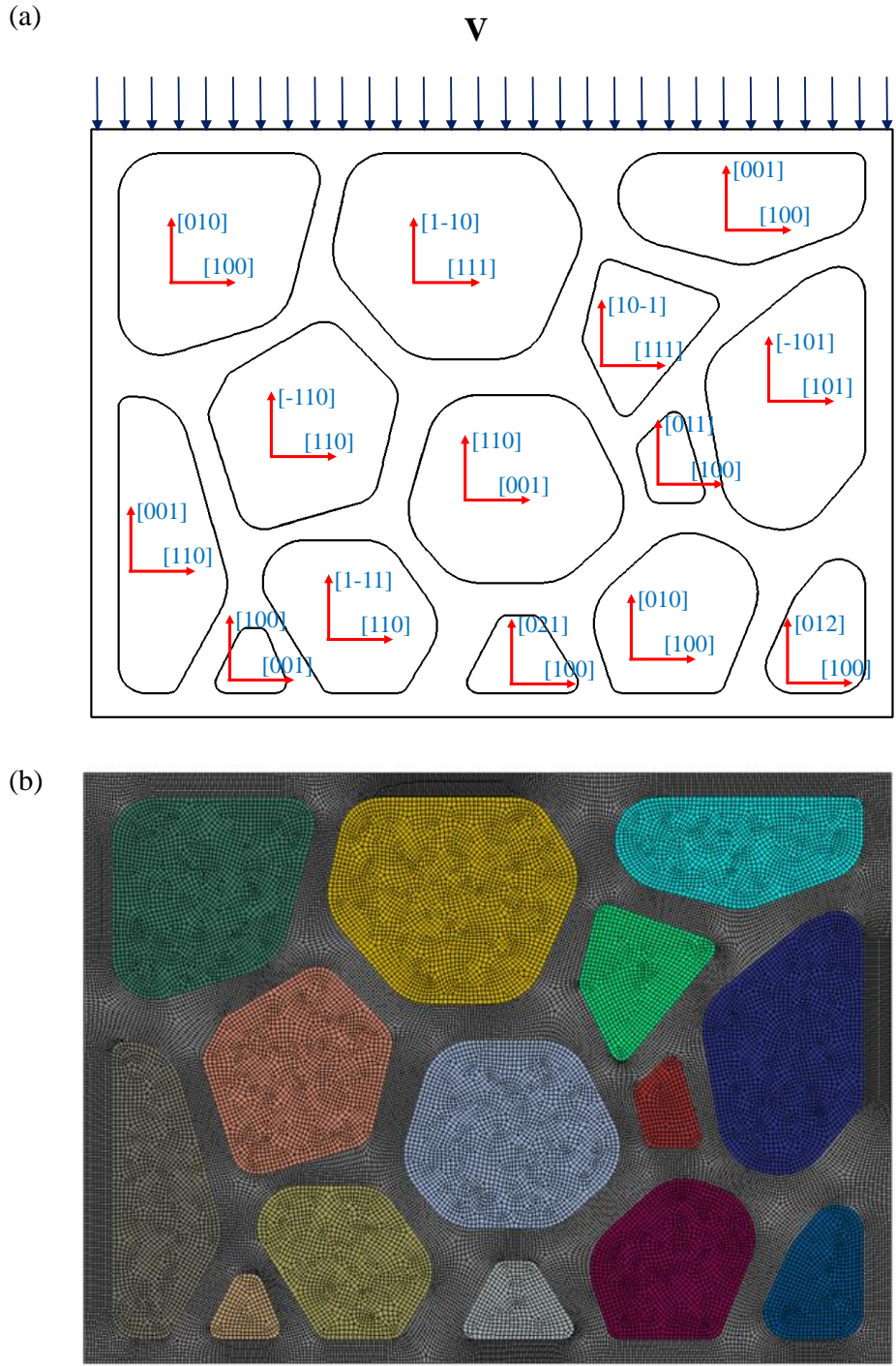

Figure 6. Meso-scale F.E simulation of a PBX aggregate (75% PETN and 25% Sylgard 182) subjected to a uniaxial compressive wave. The PBX is made of 14 PETN crystals surrounded by a Sylgard 182 matrix. (a) Orientation map of each PETN crystal; (b) F.E. mesh

As mentioned, the elasto-plastic behavior of constituent PETN crystals is described by the model given by the governing Eq. (16). For the elastic properties for the tetragonal PETN crystal we used the values reported in Sun et al. (2008) i.e. $C_{11}$ = 17.12 GPa, $C_{33}$ = 12.18 GPa, $C_{12}$ = 6.06 GPa, $C_{13}$ = 7.98 GPa, $C_{55}$ = 5.03 GPa and $C_{44}$ = 3.81 GPa (see Eq.(17) for the elastic stiffness tensor), the anisotropy coefficients involved in the yield criterion (5) are: $a_1 = 1$, $a_3 = 5.90$, $a_4 = 3.57$ and $a_5 = 6.08$ (see parameter identification in Section 3.3). The hardening law considered is of the form (see Johnson and Cook, 1983):

$$Y = \left( A + B \left( \bar{\varepsilon}^p \right)^n \right) \left( 1 + C \ln \frac{\dot{\bar{\varepsilon}}}{\dot{\bar{\varepsilon}}_0} \right)$$

where $A$= 75 MPa, $B$=59 MPa, $C$ = 0.08, $n$ = 0.233 and $\dot{\bar{\varepsilon}}_0$ = 1. For the PETN material, the heat capacity is $C$=1000 J.kg$^{-1}$.K$^{-1}$ and the Taylor–Quinney coefficient $\alpha$ =0.92. The Sylgard 182 polymer binder is assumed to be linear elastic with Young modulus E = 10 MPa, and Poisson coefficient ν = 0.48.

As shown in Figure 6(b), this PBX aggregate is meshed with 43471 hexahedral elements with reduced integration (Abaqus C3D8R solid element). The interaction between the energetic crystals and the polymer binder is assumed to follow the Coulomb friction law with a friction coefficient of 0.7. Furthermore, normal decohesion between the binder and the energetic crystal is considered when the contact normal stress is larger than 0.2 MPa, which is the tear strength of Sylgard 182. The PBX aggregate is subjected to a uniaxial compressive wave through the imposed velocity V applied at the upper surface. The applied velocity reaches 1000m/s at a time T = 0.1 μs and is kept constant for $t \in [T, T + 0.4\mu s]$. In Figure 7 is shown the rise in temperature (in K) at several times as the compressive wave travels through the energetic aggregate. Note that at a time t = 3.4 μs, the compressive wave has reached the bottom surface of the energetic aggregate. It is worth noting that the rise in temperature inside the energetic aggregate and inside individual energetic crystals is not homogenous. The results shown in Fig. 7 indicate that for locations that are far enough from the contact boundaries between the polymer binder and the energetic crystals, the rise in temperature is mainly influenced by the relative orientation of the energetic crystal with respect to the loading wave.

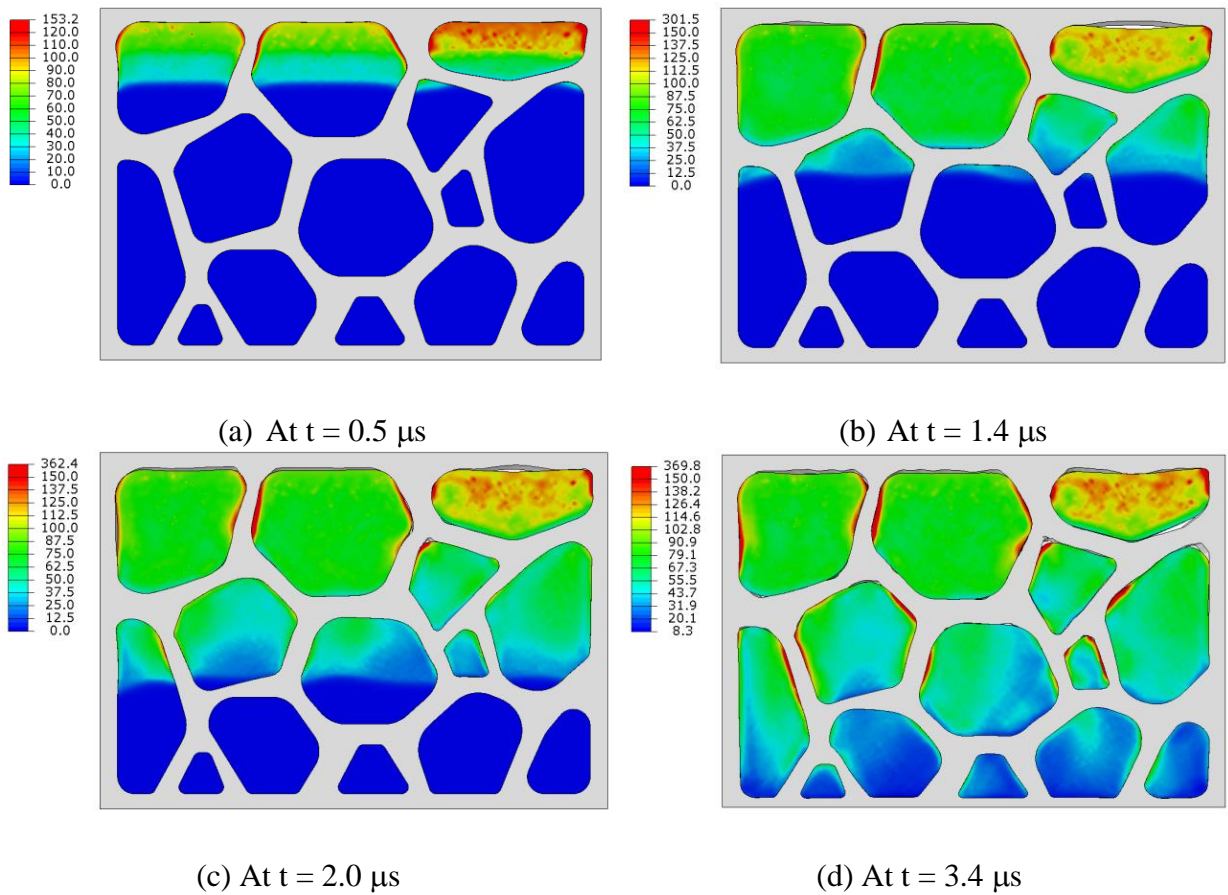

(a) At t = 0.5 μs  (b) At t = 1.4 μs

(c) At t = 2.0 μs  (d) At t = 3.4 μs

Figure 7. Rise in temperature inside a PBX aggregate consisting of 14 PETN crystals surrounded by a Sylgard182 matrix (in gray) at different times: (a) t = 0.5 μs; (b) t = 1.4 μs; (c) t = 2.0 μs and (d) t = t = 3.4 μs

As an example, in Fig. 8 is shown the evolution with time of the rise in temperature of three energetic crystals located close to the upper surface of the aggregate. The overall rise in temperature of the crystal is dictated by the plastic dissipation inside the crystal, so it is mainly influenced by its orientation (see Fig. 8 for comparison between crystal (2) and (3)). The rise in temperature experienced by the grains (1) and (2) is similar, being 71K and 75 K, respectively. However, for the grain (3) which has its [001] direction aligned with the loading direction (see Figure 6(a)), the rise in temperature is about 118 K, which is about 60% larger than in the grains (1) and (2). The interaction between the polymer matrix and the energetic crystal also causes a rise in temperature. Indeed, examination of Figure 7 shows that in each grain the maximum increase in temperature occurs at the boundary. However, the increase in temperature is strongly dependent on the orientation of the crystal-polymer boundary with respect to the loading direction. If this boundary is oriented such as to facilitate shearing, the rise in temperature will be more important (see Figure 7-8). It is also worth noting that the mesoscale simulation in which the behavior of the constituents is explicitly modeled enables to capture the decohesion that is occurring between some of the constituent crystals and the polymer matrix (see the "opening " space (shown in white) between the matrix and some crystals in Figure 7-9).

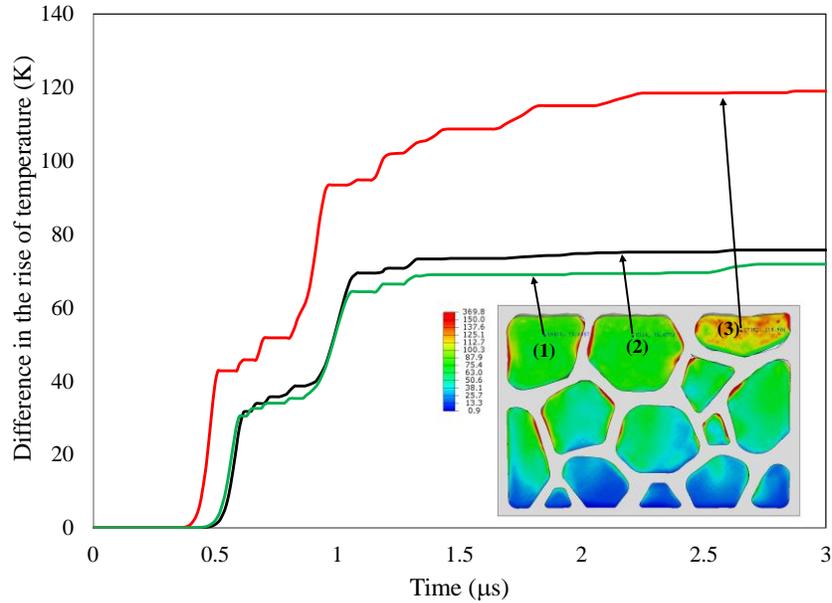

Figure 8. Rise in temperature for PETN crystals of different orientations with respect to the compressive wave (see Figure 6(a)). Note that the rise in temperature strongly depends on the crystal orientation.

Moreover, examination of Figure 7 and Figure 8 enables to assess the rise in temperature inside the PBX aggregate as the compressive wave travels through the system. When the compressive wave reaches the bottom surface, the wave is reflected. As this reflected wave travels back through the aggregate, the PETN aggregate experiences a secondary loading, which causes further changes in temperature in the constituent crystals. A comparison of the rise in temperature within the aggregate at t=3.4 μs, i.e. at a time before the wave reflects at the bottom surface ( Figure 8(d)) and at t=5.4μs when the reflected wave has passed through the aggregate (see Figure 9) shows that this secondary loading produces a non-negligible rise in temperature.

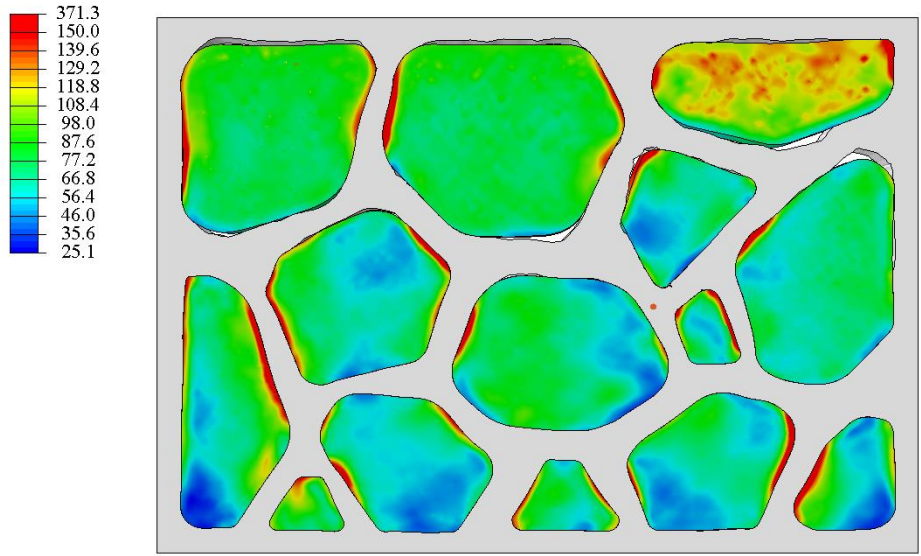

Figure 9. Rise in temperature inside a PBX aggregate made of 14 PETN crystal surrounded by a Sylgard182 matrix (in gray) as the reflected wave travel back through the aggregate (t = 5.4 μs).

## 6. Conclusions

A new anisotropic single crystal yield criterion that accounts for the specific symmetries of tetragonal crystals has been developed. This tetragonal yield criterion for energetic crystals is defined for any type of loading (fully three-dimensional stress states). It involves three independent $a_3, a_4, a_5$, which can be expressed analytically in terms of the characteristics of the tetragonal lattice structure (namely the ratio *c/a*) and the uniaxial yield stresses along four crystallographic orientations among the [100], [001], [111], [101] and [110] directions.

The capabilities of this new anisotropic yield criterion to describe the effect of loading orientation on the plastic properties of tetragonal single crystals have been demonstrated. Moreover, this new yield criterion has been applied to a PETN energetic crystal, and a very good agreement between the model predictions and uniaxial compression data was obtained. It was shown that this new yield criterion can account for the strong anisotropy of this energetic crystal. Specifically, it predicts that the maximum yield stress is along the [001] direction (**c**-axis), the ratio between the yield stress along this direction and the [100] direction (**a**-axis) being 1.85 the lowest predicted yield stress is along an orientation close to the [506] crystallographic direction, the yield stress ratio being of 0.78. In other words, for this PETN single crystal, the maximum variation in yield stress in uniaxial compression is about 134%. A new single crystal elasto-plastic model with yielding described by the new tetragonal criterion was implemented in a FE framework and further used to describe the behavior of the constituent grains in a polymer bonded explosive system subjected to dynamic compression. The FE meso-scale simulations show that the rise in temperature inside this energetic aggregate and inside each energetic crystal is not homogenous. While the temperature increase of any PETN crystal is mainly influenced by its relative orientation with respect to the loading direction (i.e. the anisotropic crystallinity), the interactions between the polymer matrix and the crystals may also contribute to severe local increases in temperature.

## References

Abaqus (2014) Abaqus - Version 6.14-1. Dassault Systemes Simulia Corp., Providence, RI

Boehler J.P. 1987 Representations for Isotropic and Anisotropic Non-Polynomial Tensor Functions. In: Boehler J.P. (eds) Applications of Tensor Functions in Solid Mechanics. International Centre for Mechanical Sciences (Courses and Lectures), vol 292. Springer, Vienna.

Boyden, F. P. and Yoffe, A.D. 1952. Ignition and growth of explosions in liquids and solids, Cambridge University Press.

Cady, H.H. and Larson, A.C., 1975. Pentaerythritol tetranitrate II: its crystal structure and transformation to PETN I; an algorithm for refinement of crystal structures with poor data. Acta Crystallographica Section B: Structural Crystallography and Crystal Chemistry, 31, pp. 1864-1869.

Cazacu, O., Revil-Baudard, B. and Chandola, N. 2019. Plasticity-Damage couplings: from single crystal to polycrystalline materials, Springer.

Clayton, J.D. and Becker, R., 2012. Elastic-plastic behavior of cyclotrimethylene trinitramine single crystals under spherical indentation: Modeling and simulation. Journal of Applied Physics, 111, pp.063512

Conroy, M., Oleynik, I.I., Zybin, S.V. and White, C.T., 2007, December. Anisotropic constitutive relationships in energetic materials: PETN and HMX. In AIP Conference Proceedings, 955, pp. 361-364.

Dick, J.J., 1984. Effect of crystal orientation on shock initiation sensitivity of pentaerythritol tetranitrate explosive. Applied Physics Letters, 44, pp. 859-861.
25